\documentclass[aps,prl,twocolumn,showpacs,superscriptaddress]{revtex4-1}
\usepackage[dvips,pdftex]{graphicx}
\usepackage{hyperref}
\usepackage{amsmath}
\usepackage{amssymb}
\usepackage{ulem}
\usepackage{color}



\begin{document}

\title{Localization in periodically modulated speckle potentials}
\author{O.S.~Vershinina $^1$,  E.A.~Kozinov $^{1,2}$, T.V.~Laptyeva$^{1,2}$, S.V.~Denisov$^{2,3}$, and M.V.~Ivanchenko $^2$}

\affiliation{ $^1$ Institute of Supercomputing Technologies, Lobachevsky State University of Nizhny Novgorod, Russia \\
$^2$Institute of Information Technologies, Mathematics and Mechanics, Lobachevsky State University of Nizhny Novgorod, Russia \\
$^3$ Department of Theoretical Physics, University of Augsburg, Germany }

\begin{abstract}
Disorder in a 1D quantum lattice induces Anderson localization of the eigenstates and drastically alters transport properties of the lattice. 
In  the original Anderson model, the addition of a periodic driving increases,  in a certain range of the driving's frequency and amplitude, 
localization length of the appearing Floquet eigenstates. 
We go beyond the uncorrelated disorder case and  address the experimentally relevant situation when spatial correlations are present in the lattice potential.
Their presence induces the creation of an effective mobility edge in the  energy spectrum of the system. 
We find that a slow driving leads to  resonant hybridization of the Floquet states, by increasing both the  participation numbers and effective widths of the states 
in the strongly localized band and decreasing values of these characteristics for the states in the quasi-extended band.
Strong driving homogenizes  the bands, so that the Floquet states loose compactness and tend to be spatially smeared.  
In the basis of the stationary Hamiltonian, these states retain localization in terms of participation number
but become de-localized and spectrum-wide in term of their effective widths. Signatures of thermalization are also observed.
\end{abstract}

\maketitle

Anderson localization in disordered systems is a fundamental phenomenon that is still posing new puzzles and bringing new surprises \cite{Kramer1993,Evers2008,fifty}. 
The original problem of non-interacting quantum particles \cite{And58} was studied thoroughly and has been placed 
in a broad context, resulting in experimental observations of the localization 
with matter \cite{Billy08,Roati08,Kondov11,Jendr2011}, electromagnetic \cite{Optics}, and acoustic  waves \cite{Hu08}. 

The effect of periodic modulations on the localization also received considerable attention. 
It was found that the localization length increases under the low frequency 
driving (though non-monotonously with the driving  amplitude) and decreases in the opposite limit of the fast driving \cite{Martinez2006}. 
The increase of the localization length was attributed to the induced interaction between the particle path channels, with
those characterized by  weakest localization properties making a dominant contribution. 
In contrast, the high-frequency driving diminishes time-averaged hopping amplitudes 
\cite{Martinez2006,Holthaus1995,Holthaus1996} and enhance the localization, an effect 
reminiscent of the dynamic localization \cite{Dunlap1986,Hanggi1991}. 
Recently, it has been shown that the multi-frequency driving can substantially 
increase the localization length  \cite{Flach2016}, 
and the complete de-localization can be achieved with driven quasi-periodic potentials \cite{Flach2014}.

The existing results, however, address  the original Anderson set-up, 
with on-site  energies being random and {\it uncorrelated} variables. 
At the same time, the presence of {\it correlations} is inherent to the optical speckle potentials, 
used in the experiments with atomic Bose-Einstein condensates \cite{Modugno2006}.
Importantly, a finite correlation length leads to emergence of an effective mobility 
edge separating the bands with localization lengths  differing by orders of magnitude \cite{Lugan2009,Modugno2010,Giacomelli2014}.

Application  of periodic modulations to a system with correlated disorder
evokes a set of intriguing questions. For example, will the driving modify 
quasi-extended states in a way different from it does in the case of uncorrelated disorder? 
Can the driving induce a coupling across the mobility edge? Can the co-existence of the strongly 
localized and quasi-extended states change the localization picture known for the modulated system with the uncorrelated disorder?

In this Letter we study the fate of localization in periodically and 
slowly driven 1D speckle potential. We find that the driving creates  a more intricate picture than in the case of uncorrelated disorder. 
Increase of the driving amplitude leads to the appearance of  complex structure in the set of the induced  Floquet states. 
Participation numbers grow for strongly localized states and decrease for quasi-extended. 
At the same time both kinds of states loose compactness and become sparse and weakly localized. 
Similar transformation occurs in the basis of the stationary Hamiltonian, where the Floquet states 
become de-localized in terms of the effective width. These changes are underpinned by resonant hybridization of Floquet states, 
as our numerical results and perturbative arguments indicate. Strong driving leads to a mixing of states, by inducing a homogeneity of their
localization properties and brings traits of thermalization, as supported by their center of mass and average energy statistics.

\begin{figure}[t]
\begin{center}
\includegraphics[width=0.9\columnwidth,keepaspectratio,clip]{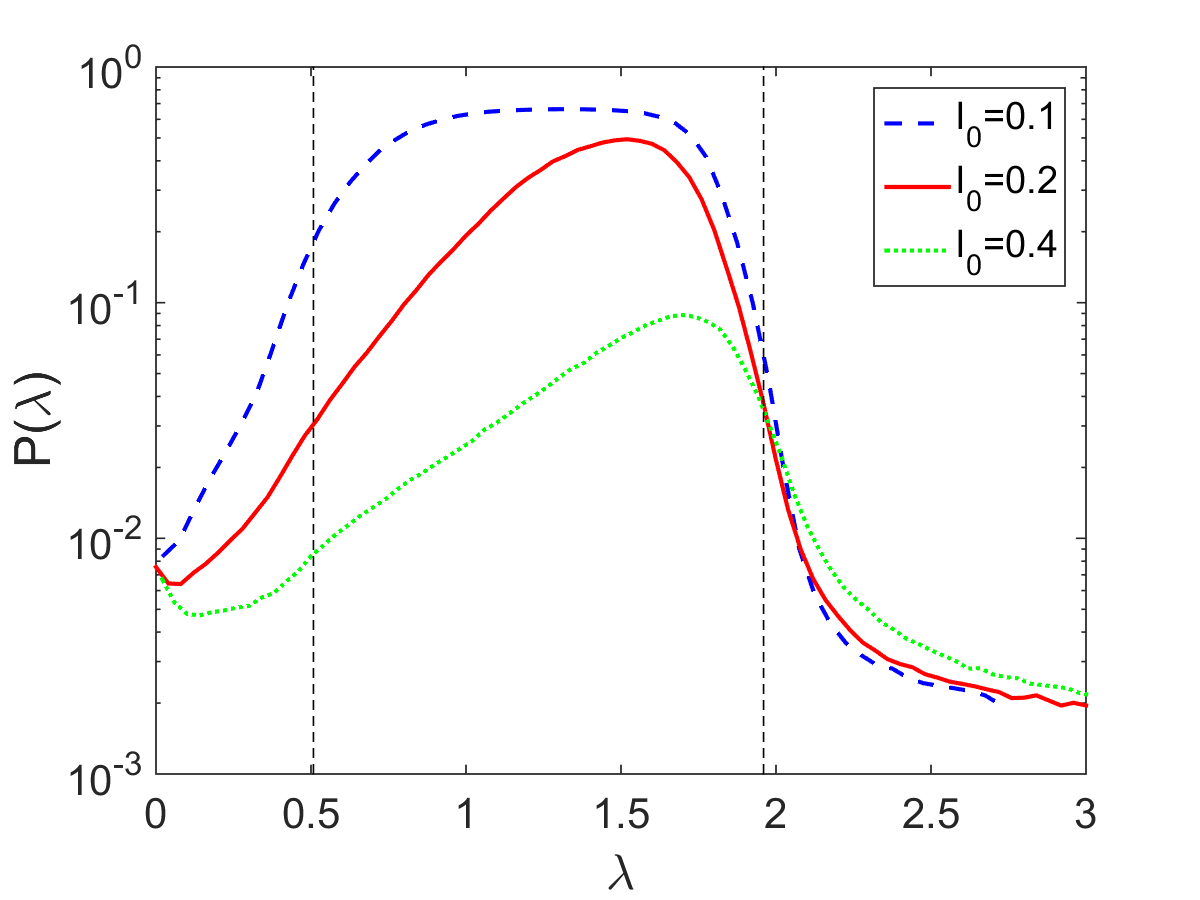}
\caption{Localization in the stationary lattice with correlated disorder: 
average participation number of the eigenstates [normalized by the system size, $P=1/(N\sum|\psi_j|^4)$]
as functions of energy, $\lambda$, for different values of the disorder strength $I_0$. 
Vertical dashed lines indicate the band of quasi-extended states: 
effective mobility edge (left) and the border with ``semi-classical'' strong localization (right).} 
\label{fig:0}
\end{center}
\end{figure}

\begin{figure*}[t]
\begin{center}
(a) \includegraphics[width=0.6\columnwidth,keepaspectratio,clip]{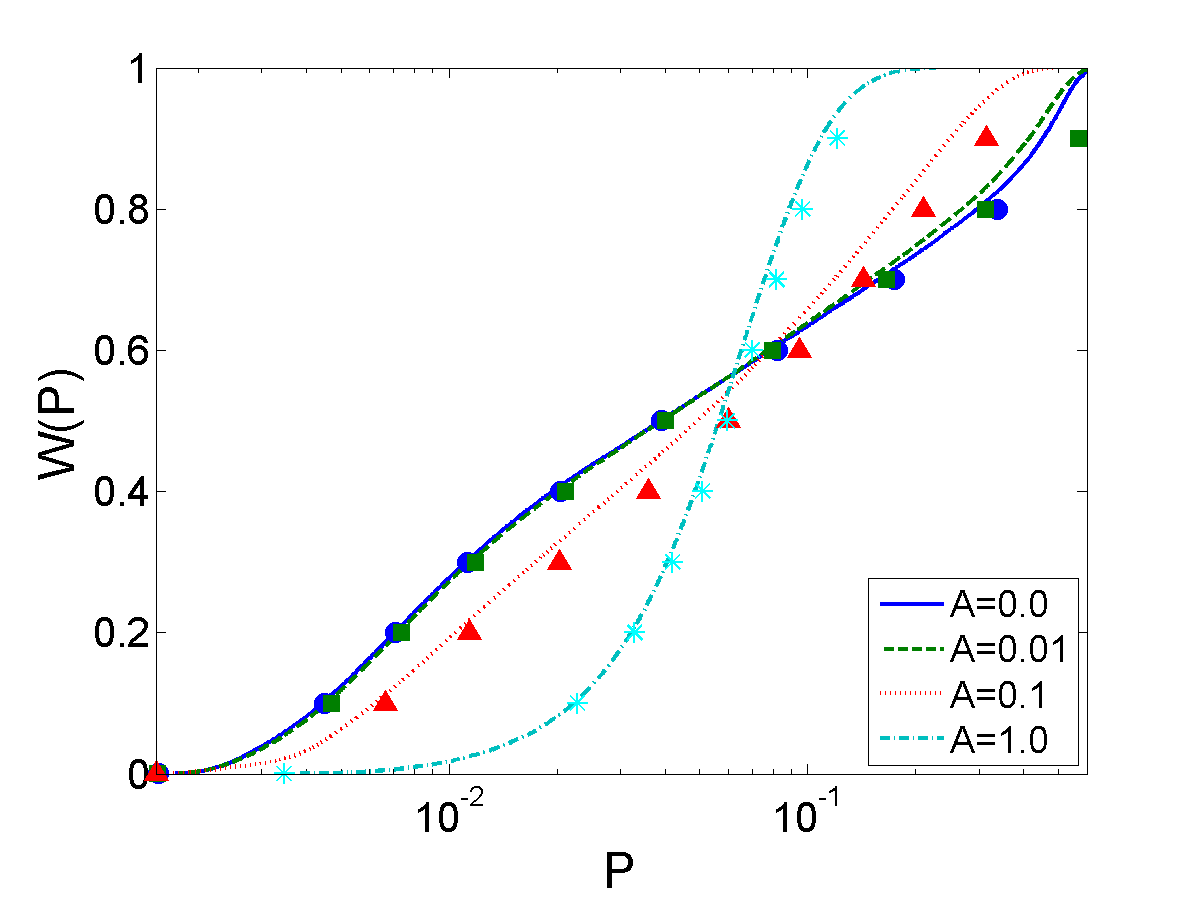}
(b) \includegraphics[width=0.6\columnwidth,keepaspectratio,clip]{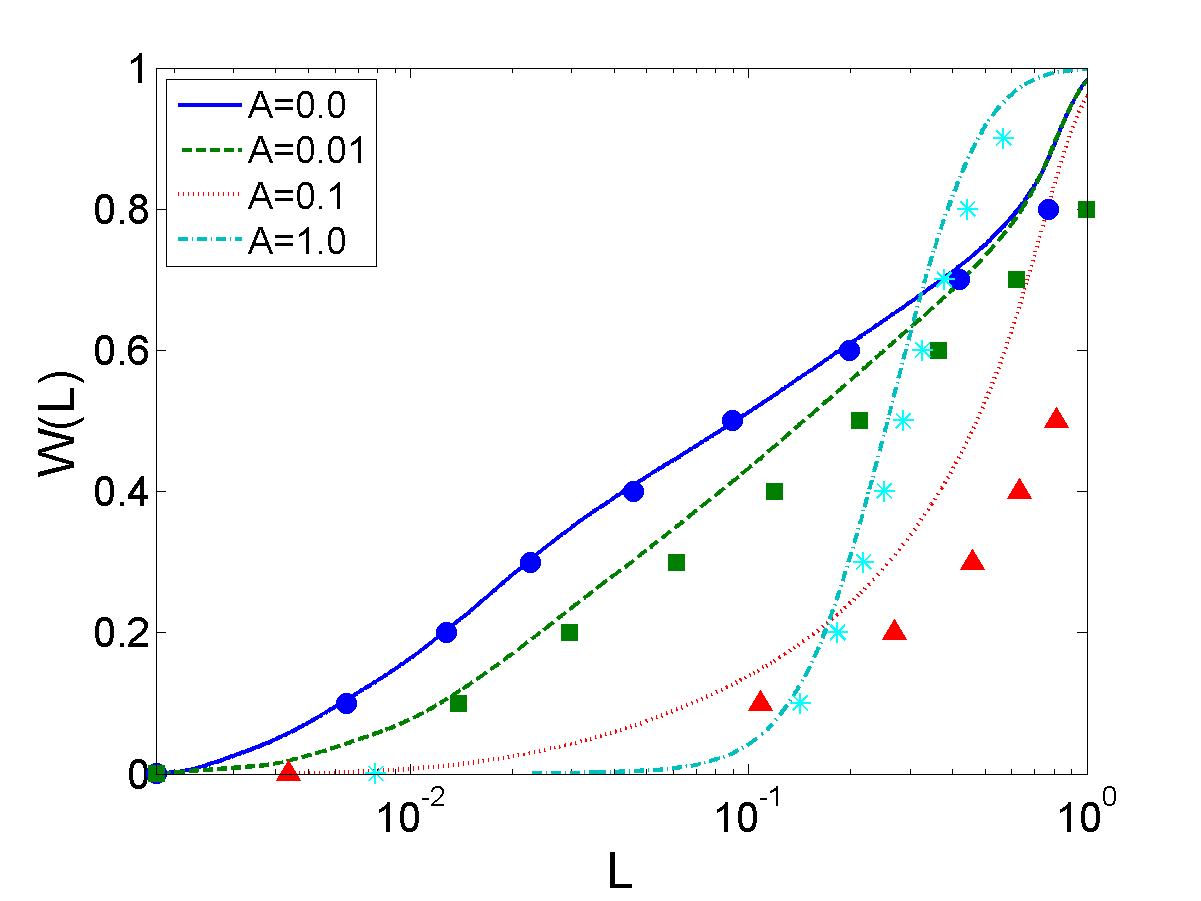}
(c) \includegraphics[width=0.6\columnwidth,keepaspectratio,clip]{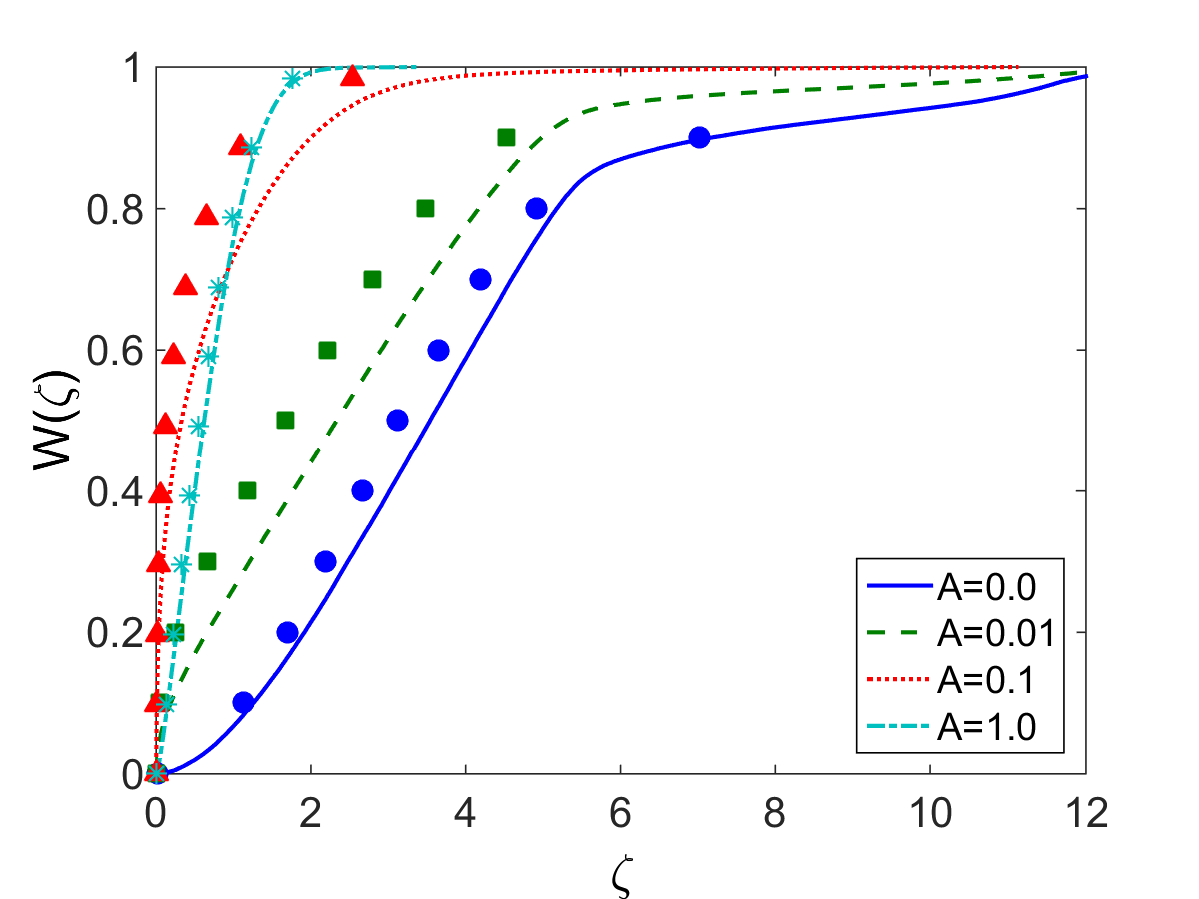}\\
(d) \includegraphics[width=0.6\columnwidth,keepaspectratio,clip]{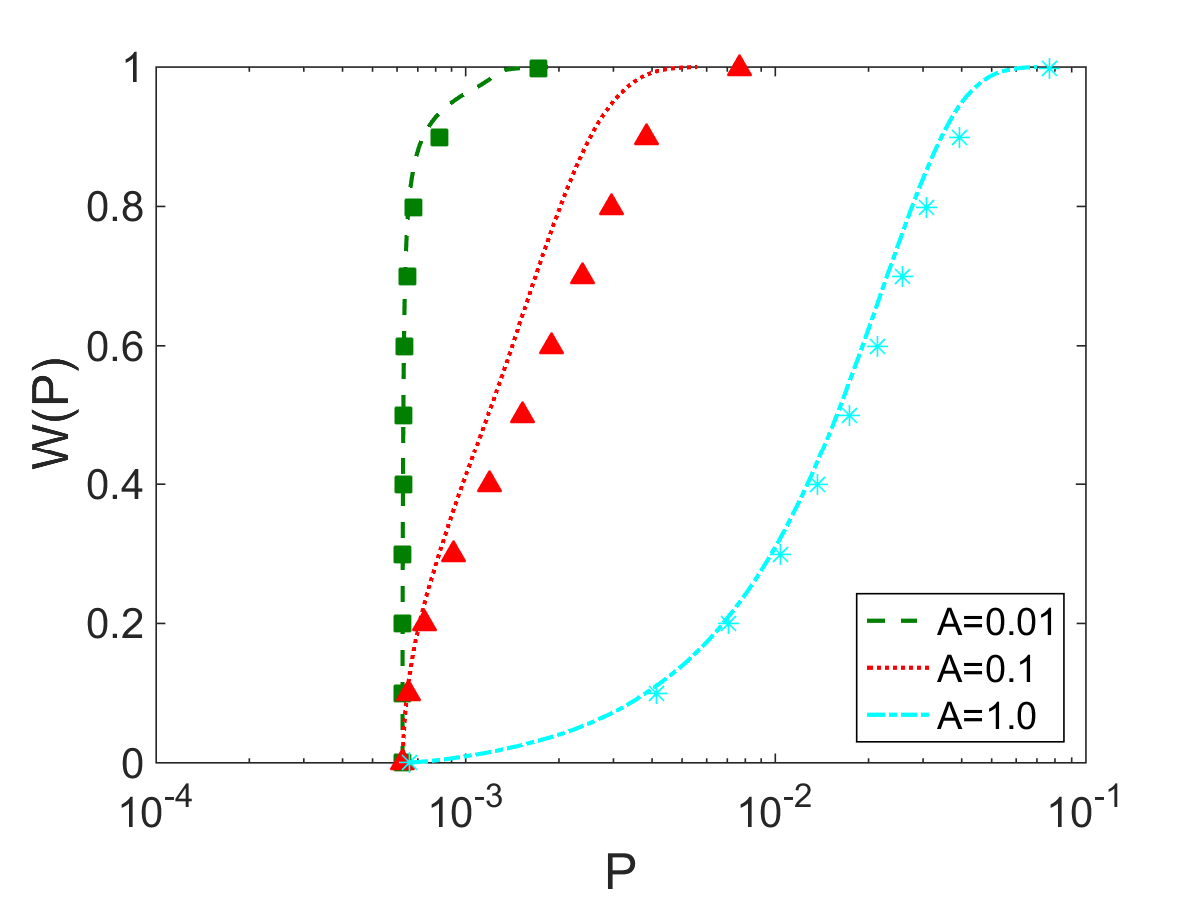}
(e) \includegraphics[width=0.6\columnwidth,keepaspectratio,clip]{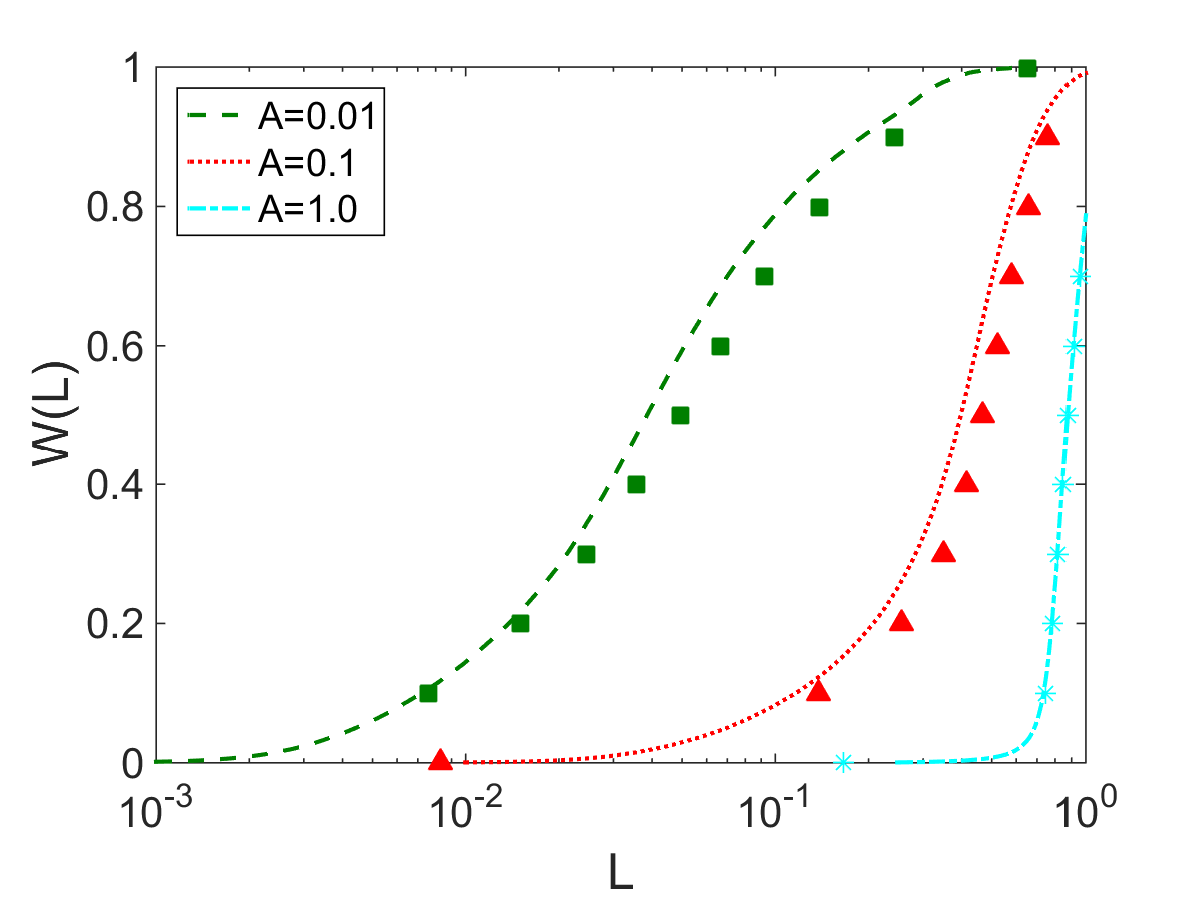}
(f) \includegraphics[width=0.6\columnwidth,keepaspectratio,clip]{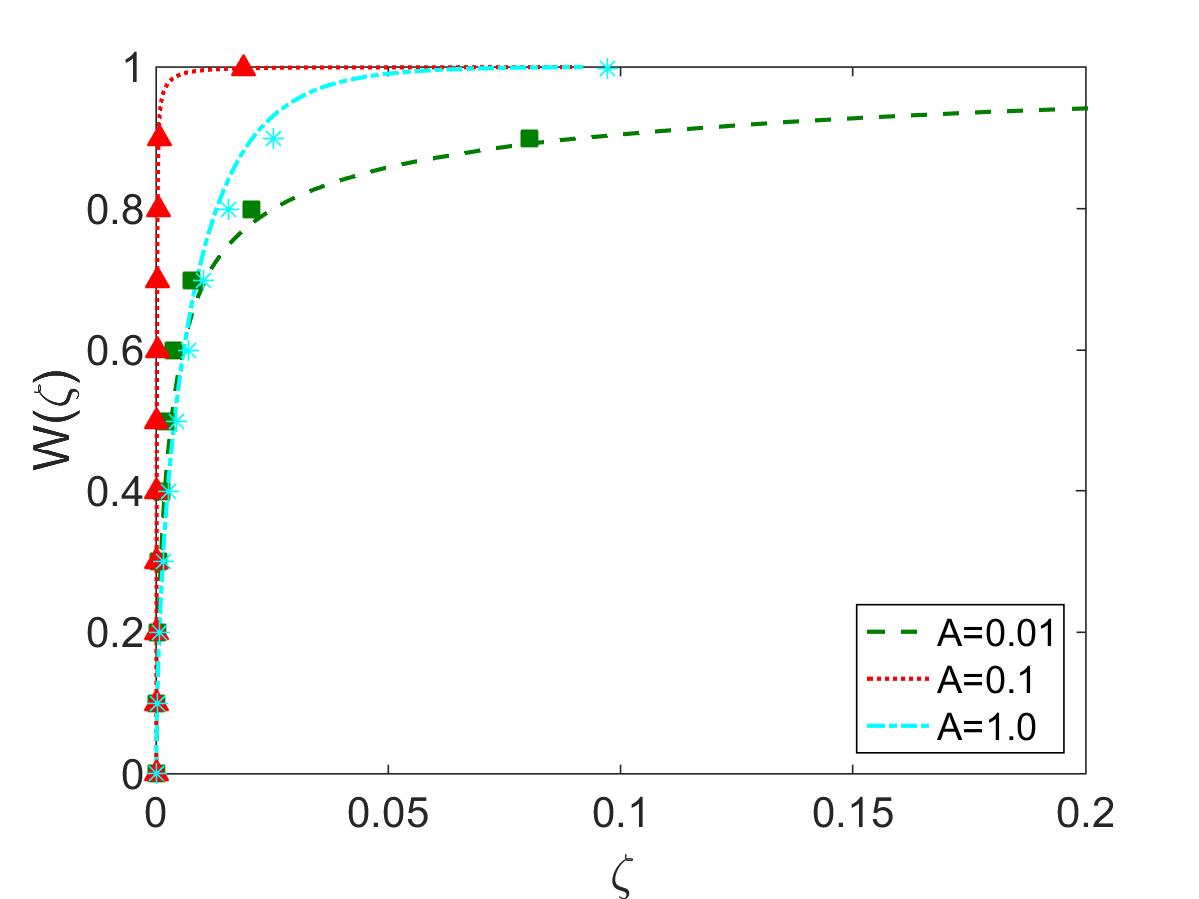}

\caption{Localization properties of Floquet states as functions of  driving amplitude $A$: 
integral distributions of (a) normalized participation number $P$, (b) normalized effective width $L$, 
and (c) compactness index, $\zeta$; (e,c,f) -- the same in the basis of the stationary Hamiltonian. 
Here $\omega=0.5, I_0=0.2$. 
Lines show the results obtained for $N=1600, D=400$ and averaged  over $N_r=100$ disorder realizations.
Symbols of respective color correspond to single disorder realizations for $N=6400, D=1600$ 
($A=0$, `$\bullet$', $A=0.01$, `$\blacksquare$', $A=0.1$, `$\blacktriangle$', $A=1.0$, `*'), 
and the scaling $4\times P$ (a,e), $4\times L$ (b) and $16\times\zeta$ (f) is implemented (see the main text for more details).}
\label{fig:2}
\end{center}
\end{figure*}

We consider the dynamics of a quantum particle in a speckle potential described by a periodically-modulated  tight-binding Hamiltonian,
\begin{equation}
\label{eq:1}
H(t)=\sum_j \epsilon_j\left(1+A \cos\omega t\right) b_j^{\dagger}b_j+\frac{1}{2}\left(b_{j}^{\dagger} b_{j+1} + b_{j+1}^{\dagger} b_{j}\right), 
\end{equation}
where $b_j$ and $b_j^{\dagger}$ are the annihilation and creation operators of a particle at a site $j$, $j = 1,...,N$, 
and $\epsilon_j>0$ are correlated random on-site energies of speckle potential \cite{Modugno2006}.  $A$ and $\omega$ are the  amplitude and frequency of the modulations.

On-site energies
are defined by the experimental set-up, where electric field $\mathcal{E}(x)$ (created by a laser speckle) 
is,  to a very good approximation, a realization of a complex Gaussian variable \cite{Modugno2006}. 
Quantum particles experience the potential corresponding to the field intensity, $\epsilon(x)=|\mathcal{E}(x)|^2$, 
or $\epsilon_j\equiv\epsilon(x_j)$ in the tight-binding approximation (\ref{eq:1}). 
The energy distribution  across the speckle pattern follows Rayleigh distribution, $W(\epsilon)=\exp(−\epsilon/I_0)/I_0$, 
where $I_0$ is the mean intensity. The intensity autocorrelation function is  $\left\langle \epsilon(x)\epsilon(0)\right\rangle=I_0[1+{\rm{sinc}}^2(D x)]$, 
where $D$ is the aperture. The (auto) correlation length $\xi$ of the speckle potential is defined through the equation ${\rm{sinc}}^2(D\xi/2)=1/2$, 
which gives the width at the half maximum of the autocorrelation function. It is related to the
aperture width as $\xi=0.88/D$.

In the Fock basis, $|j\rangle\equiv b_j^\dagger|0\rangle$, and with $m=\hbar=1$, the corresponding Schr\"{o}dinger equation assumes the form
\begin{equation}
\label{eq:2}
i\dot{\psi}_j=\epsilon_j\left(1+A \cos\omega t\right) \psi_{j}+\frac{1}{2}\left(\psi_{j+1} + \psi_{j-1}\right).
\end{equation}

\begin{figure*}[t!!]
\begin{center}
(a)\includegraphics[width=0.9\columnwidth,keepaspectratio,clip]{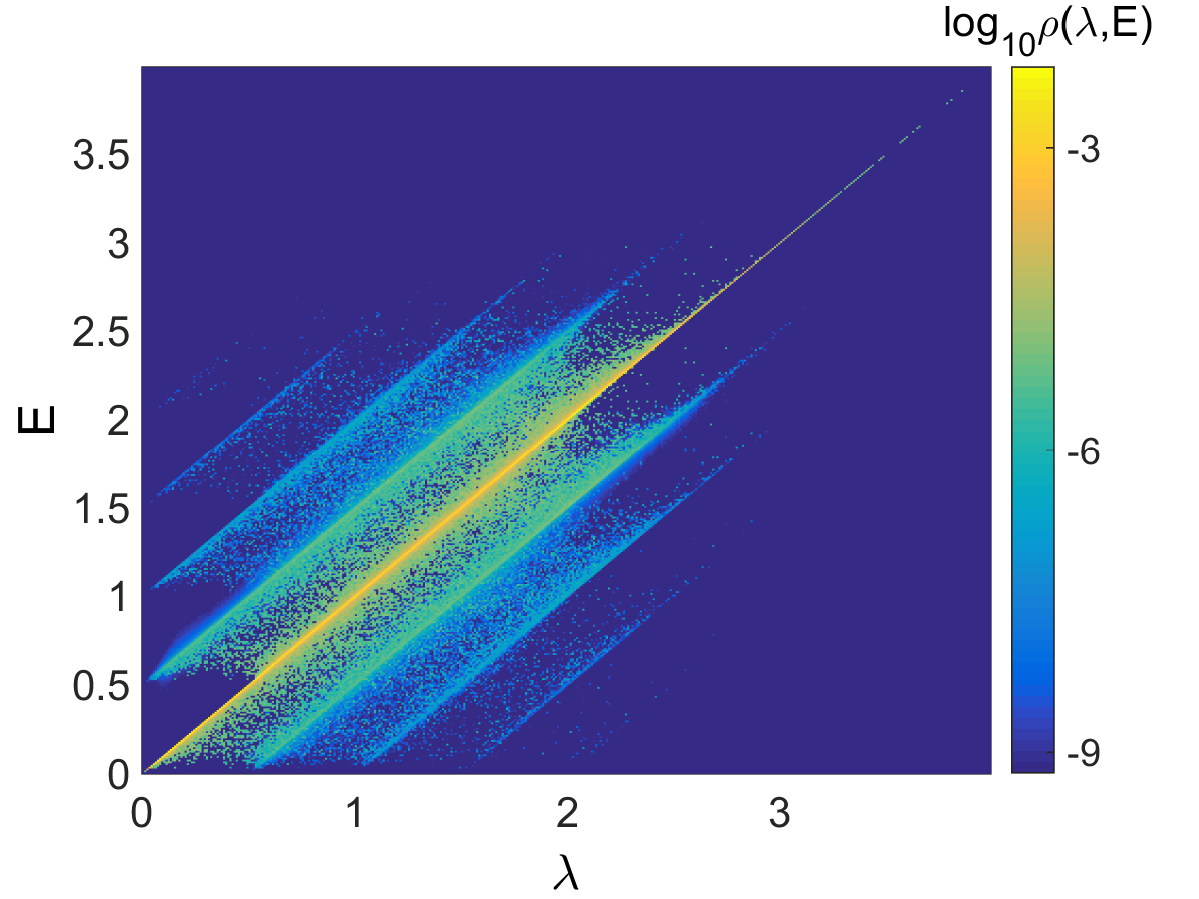} 
(b)\includegraphics[width=0.9\columnwidth,keepaspectratio,clip]{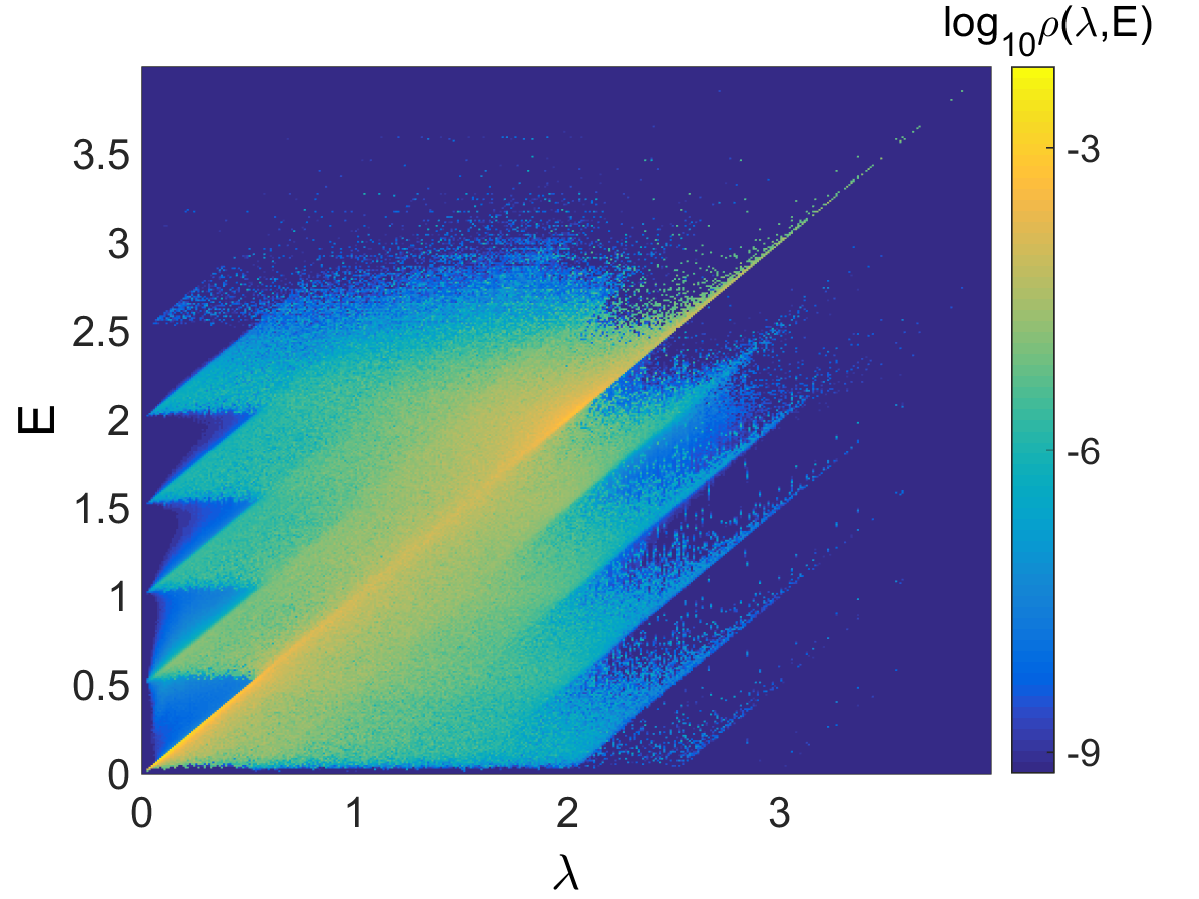}\\
(c)\includegraphics[width=0.9\columnwidth,keepaspectratio,clip]{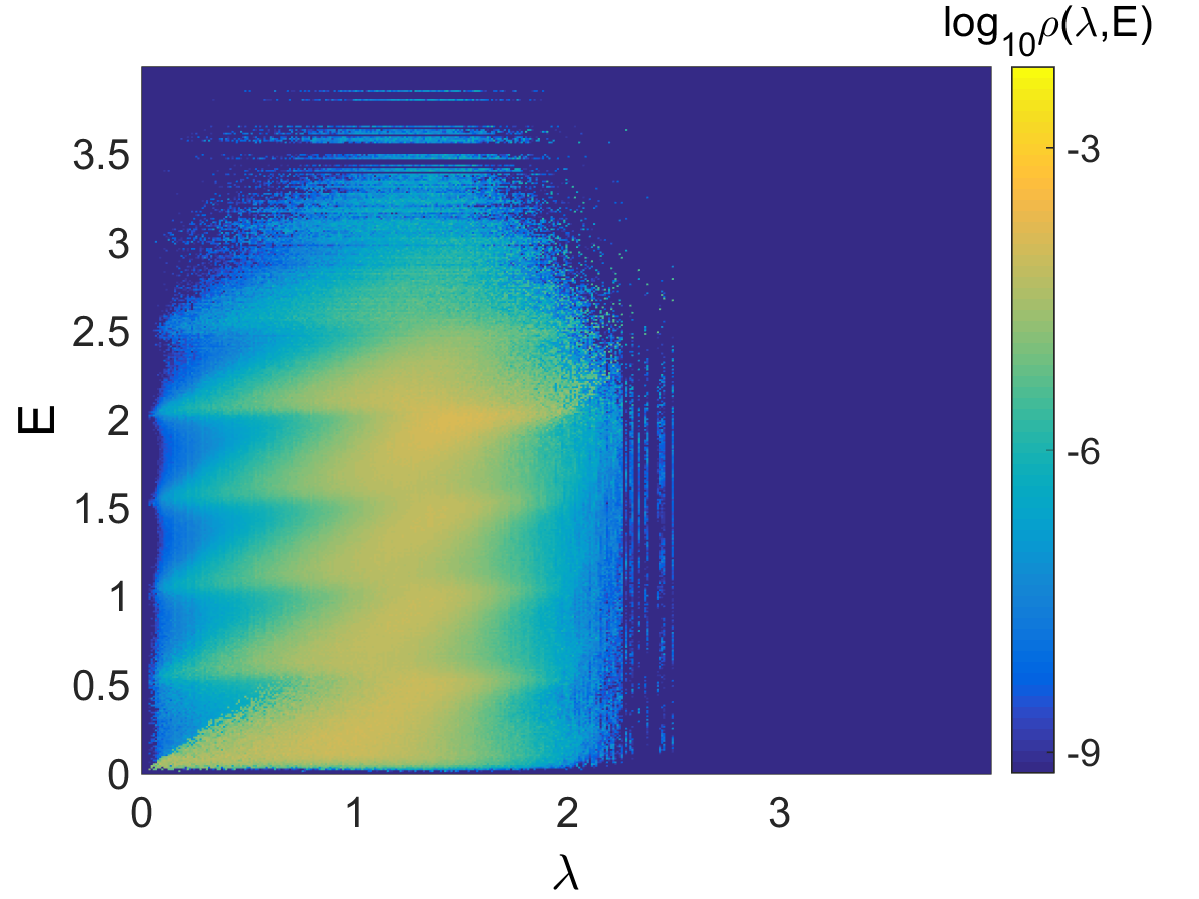}
(d)\includegraphics[width=0.9\columnwidth,keepaspectratio,clip]{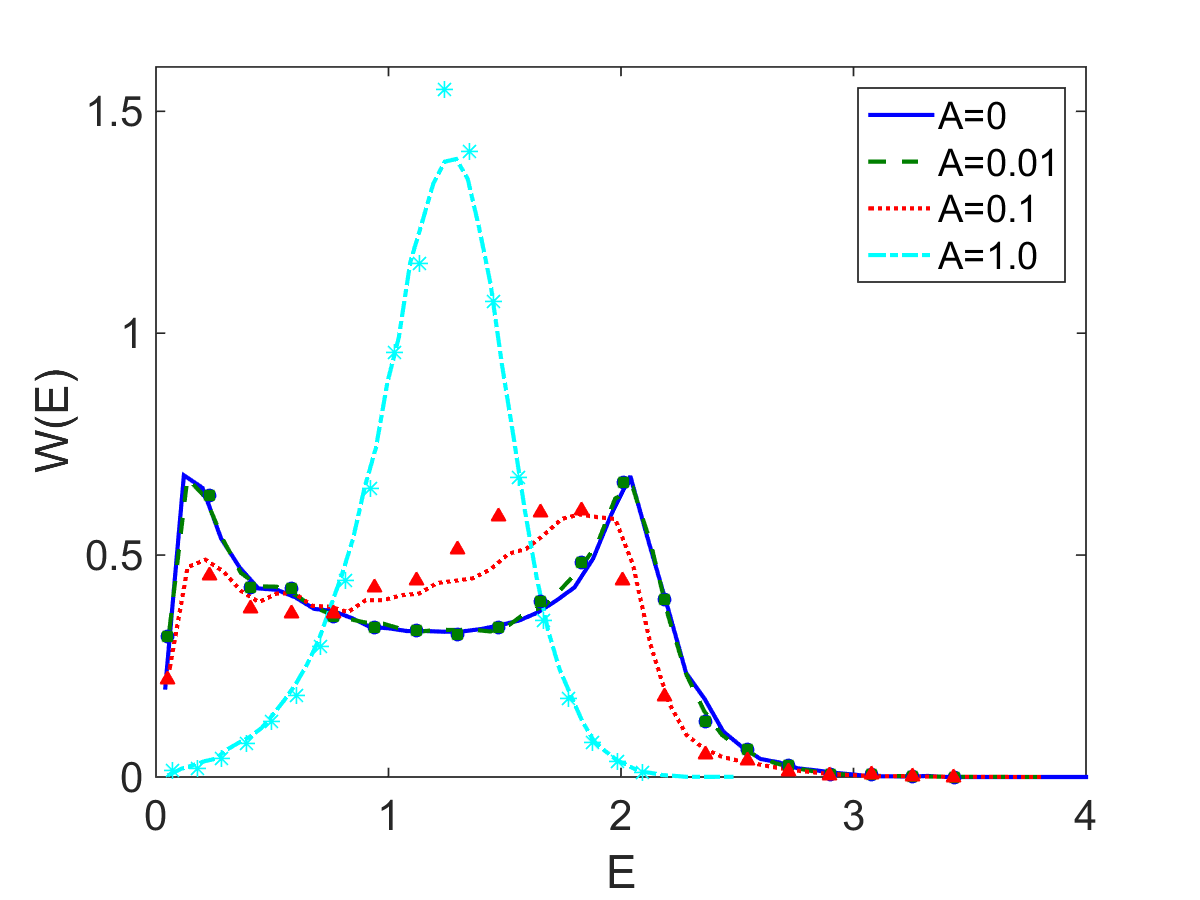}
\caption{The probability density function $\rho(\lambda,E)$ obtained by expanding Floquet states  in the basis of the stationary Hamiltonian, 
where $E$ is the energy of Floquet states averaged over the period of the modulations  and $\lambda$ is the eigen-energy of the eigen-states of the stationary Hamiltonian.
The distribution are calculated for different amplitude of driving, (a) $A=0.01, (b) 0.1$ and (c) $1.0$. 
Here $I_0=0.2, \omega=0.5$, $N=1600$, $D=400$, and $N_r=100$. 
(d) Probability density distributions for the Floquet state specific energy (lines correspond to $N=1600$, $D=400$, $N_r=100$, symbols to $N=6400$, $D=1600$, $N_r=1$). Emergence and sharpening of the maximum indicates homogenization of the states  and thermalization of the system. } 
\label{fig:1}
\end{center}
\end{figure*}

The key property of the stationary Hamiltonian, $A=0$, is the presence of an effective mobility edge in 
its spectrum, estimated as $\lambda^M=I_0+\left(0.88\pi/\xi\right)^2/2$  \cite{Lugan2009,Modugno2010}. It
separates strongly and quasi-extended states, Fig. \ref{fig:1}(a). 
The localization length of the latter is one or two orders of magnitude larger
than that of the former so that in finite systems the  quasi-extended states
can be treated as effectively de-localized \cite{Giacomelli2014}. 
At higher energies, the strong localization is restored as a ``semi-classical'' localization. 

Dynamics of the driven system can be fully evaluated in terms  of the so-called ``Floquet eigenstates'' \cite{shirley,sambe, grifoni}. 
They are the eigenstates of the unitary Floquet propagator $U_T = \mathcal{T} \exp\left[-\frac{i}{\hbar} \int_0^T H(\tau) \mathrm{d}\tau\right]$, 
where $\mathcal{T}$ is the time-ordering operator. The particular structure of
the Floquet propagator, and thus the localization properties of Floquet states, depend on the parameters of the driving. This dependence -- in the context of localization --
is the main objective of our study.

The Floquet states are found by numerically integrating Eq.~(\ref{eq:2}) over one period of driving,  by using a 
high-precision Magnus-Chebyshev scheme \cite{Laptyeva2015}  with 
$200$ integration steps per period.  We choose $N/D=4$ so that the correlation length $\xi\approx3.5$ 
and make the core simulations for $N=1600\approx450\xi$. The disorder strength is set to $I_0=0.2$
so that  $\lambda^M \approx 0.5$, ensuring a substantial part of the states of both kinds; see Fig.\ref{fig:0}(a). 
Finally, zero boundary conditions are assumed.

Localization properties are quantified by the standard measures used in the field. 
Participation number is normalized by the system size, $P=1/\left(N\sum|\psi_j|^4\right)$, 
so that it manifests the effective ratio of non-zero components in a Floquet vector. 
Their second moment, $m_2=\sum\left(j-\langle j \rangle\right)^2|\psi_j|^2$, where $\langle j \rangle=\sum j|\psi_j|^2$, 
defines an effective width, $L=\sqrt{12m_2}/N$ (also normalized matching an effective uniform brick). 
Finally, we calculate compactness index, $\zeta=(P N)^2/m_2$, which ranges from $\zeta\sim 0$ for sparsely 
populated states (a few effectively non-zero well-separated elements) to $\zeta\sim3$ for random sets to $\zeta=12$ 
for an ideally compact state (uniformly populated neighbor elements). 
These quantities are averaged over a period of driving \cite{Note1}. 
Note, that for a localized state, which length does not scale with the system size, one has $P, L\sim 1/N$, while $\zeta$ will not change.

Numerical results reveal two opposite trends in the localization of Floquet states (depending on their positions with respect to the mobility edge)
when the modulation amplitude increases, $A \gg 0.01$; see Fig.\ref{fig:2}. 
Participation number increases for the strongly localized states and decreases for the quasi-extended ones, 
almost equating them when the driving is strong, $A=1$, Fig.\ref{fig:2}(a). Similar scenario is observed 
for the effective widths of the states although strong driving settles it at the values much greater than participation numbers, see Fig.\ref{fig:2}(b). 

The emerging complex structure of Floquet states is manifested in the compactness index, which rapidly  drops below $\zeta=1$ for both kinds of Floquet states, Fig.\ref{fig:2}(c). 
This reflects the hybridization of states and indicates development of the highly fragmented spatial structure. 
Local outbursts in the site population numbers, 
scattered over the lattice, give an ultimate image of Floquet states 
in the direct basis under strong driving. 
Numerics for a larger system size, $N=6400, D=1600$,
reveals qualitatively the same results and a fair quantitative match, when $P$ and $L$ are scaled with the system size as described in the above, see Fig.\ref{fig:2}(a,b,c).

Further insight can be obtained by using the basis of the stationary Hamiltonian, $\{\mathbf{z}^{(l)}\}$, such that $H\mathbf{z}^{(l)}=\lambda_l \mathbf{z}^{(l)}$, $A=0$. 
By using expansion $\boldsymbol{\psi}=\sum_l \varphi_l \mathbf{z}^{(l)}$,   Eq.(\ref{eq:2}) can be rewritten as
\begin{equation}
\label{eq:3}
i\dot{\varphi}_k=\lambda_k\varphi_k+A\cos{\omega t} \sum_{k'} J_{k,k'}\varphi_{k'},
\end{equation}
where $J_{k,k'}=\sum_j \epsilon_j z^{(k)}_j z^{(k')}_j$ are the driving-induced overlap coefficients, and the stationary states are numbered according to the eigenvalues, $\lambda_k$, in increasing order. 

Localization in this basis can also be characterized by $P, L$, and $\zeta$.  For zero driving the Floquet states coincide with the stationary basis, and are, therefore, 
compact: $P=1$, while $m_2$, $L$ and $\zeta$ are ill-defined. As the driving amplitude increases, we observe that participation number grows for all states, along 
with the effective widths. Notably, the former remains much less than one, indicating that the Floquet states expand over a small subpart of the whole basis 
only, Fig.\ref{fig:2}(d). At the same time, the latter becomes of the order of the system size, manifesting hybridization between the states from different 
parts of the spectrum, Fig.\ref{fig:2}(e). Remarkably, unlike in the direct basis, the normalized effective width, $L$, does not decrease as the system size grows, 
while the normalized participation number does, $P\sim 1/N$, see results for  $N=6400, D=1600$ presneted with Fig.\ref{fig:2}(d,e). Accordingly, most of the Floquet states are 'smeared' 
over the whole energy spectrum;  in that sense they are de-localized, although tey are remaining sparse in terms of the compactness index $\zeta\sim1/N^2$, see Fig.\ref{fig:2}(f).  

To explain these findings, we analyze the Floquet states $F_k^{(l)}(t)$, 
\begin{equation}
\label{eq:4}
F_k^{(l)}(t)=\Phi_k^{(l)}(t)\exp(-i s_l t),
\end{equation}
where $\Phi_k^{(l)}(t)$ are $T=2\pi/\omega$-periodic functions, eigenstates of the Floquet propagator, 
and $s_l$ are quasi-energies. Obviously, in the absence of driving, $A=0$, 
the Floquet states correspond to the stationary states, $\Phi^{(l)}_k=\delta_{k,l}$, and the quasi-energies are given by the respective eigenvalues (energies), $s_l=\lambda_l$.

Next we expand the Floquet states  in the basis of the eigenstates of the stationary Hamiltonian, Eq.~(\ref{eq:3})
In this picture, the Floquet states  are results of driving-induced  cross-breeding  between the eigenstates. 
By using the standard perturbation theory, we  expand $\Phi_k(t)$ into  Fourier series and express Floquet states as
\begin{equation}
\label{eq:6}
F_k(t)=\sum_{n}\phi_k^{(n)}\exp(-i n \omega t)\exp(-i s t).
\end{equation}
Substituting (\ref{eq:6}) into (\ref{eq:3}), we get
\begin{equation}
\label{eq:7}
s\phi_k^{(n)}=(\lambda_k-n\omega)\phi_k^{(n)}+\frac{1}{2}A\sum_{k'} J_{k,k'}\left(\phi_{k'}^{(n-1)}+\phi_{k'}^{(n+1)}\right).
\end{equation}

Taking the $l$-th Floquet state for $A=0$ as a zero-order approximation, 
$\phi_{k}^{(0)}=\delta_{k,l}$ and $s=\lambda_{l}$, for the first order we obtain  non-zero amplitudes 
\begin{equation}
\label{eq:9}
\phi_k^{(\pm1)}=\frac{A J_{k,l}}{2(\lambda_{l}-\lambda_k\pm\omega)}.
\end{equation}
The net norm of the first order corrections is, therefore,
\begin{equation}
\label{eq:10}
\|\Delta\phi^{(l)}\|=\frac{1}{4}A^2\sum_k\frac{J_{k,l}^2}{(\lambda_{l}-\lambda_k\pm\omega)^2},
\end{equation}
and provides with the necessary validity condition $\|\Delta\phi^{(l)}\|\ll1$.

Note that the higher order resonances will also be induced in chain, until the multiple of the driving frequency stays within the spectrum bound.
Clearly, the degree of hybridization between two Floquet states depends on both the overlap coefficient between the states, $J_{k,l}$, 
and the distance to the resonance, $|\lambda_k-\lambda_{l}|\approx n\omega$, $n\in\mathbb N$. 
To secure the latter possibility, the driving has to be {\it slow} 
so that its frequency does not exceed the characteristic spectrum width. 
Furthermore, it can be expected that most of the strongly localized states would not interact significantly, 
as both the near-resonance and overlap conditions would rarely be satisfied at the same time. 

A localized state occupies a certain localization volume $V_L\ll N$,  and normalization requires that the non-zero elements scale as  $|z_l|\sim V_L^{-1/2}$. 
We assume that the signatures of the eigenstates elements $z^{(k)}_j$ and $z^{(l)}_j$ are not correlated, 
and obtain $|J_{k,l}|\sim V_L^{-1/2} \Theta(k,l)$, where $\Theta(k,l)=1$ if the eigenstates overlap, and $0$ otherwise \cite{Note2}. 
As the probability of the latter scales as $\sim V_L/N$, we finally obtain $|J_{k,l}|\sim V_L^{1/2}/N$. 
In turn, the contribution of the frequency denominator can be estimated as follows. Since the density of states scales 
linearly with $N$, the eigenstate spacing behaves as $\sim 1/N$, and the squared resonance mismatch in the denominator yields $\sim 1/N^2$. 
Thus we arrive at $\|\Delta\phi^{(l)}\|\sim A^2 V_L$. It indicates that the first corrections due to driving-induced interactions between strongly localized 
states remain finite ( even in the limit $N\rightarrow\infty$), which is consistent with the moderate influence of driving, found in 
the case of the Anderson (uncorrelated) disorder  \cite{Martinez2006,Holthaus1995,Holthaus1996}.  


The presence of effective mobility edge and quasi-extended band due to correlations in disorder,
enriches the picture. Indeed, many quasi-extended modes would fall within the same localization 
volume $V_E\gg V_L$ and interact efficiently. Moreover, they may couple across the mobility edge to the 
strongly localized ones, and affect the latter. Repeating the above arguments, for the overlap integrals 
between quasi-extended states or between such a state and a strongly localized one, we 
obtain $\|\Delta\phi^{(l)}\|\sim A^2 V_E$. Given $V_E/V_L\sim10^1\ldots10^2$ (cf. Fig.\ref{fig:0}), it follows that reshaping of Floquet states is dominated by the coupling to and between quasi-expended modes.     

Now one can explain the observed changes in the localization properties in both bands. 
It is straightforward  to see that in a 
hybrid state, composed of a localized and a quasi-extended ones, 
most contribution to participation number steams from the former due to large local amplitudes. 
On the contrary, most contribution to the effective width is produced by the latter. 
Accordingly, the majority of the Floquet states eventually acquires relatively low participation numbers and relatively high effective widths, thus becoming sparse, see 
Fig.\ref{fig:2}(a-c). Still, in the thermodynamic limit $N\rightarrow\infty$, such hybridization cannot lead to a complete de-localization 
in the direct basis as even quasi-extended states still have finite, albeit huge, localization length \cite{Lugan2009,Modugno2010,Giacomelli2014}.

Moreover, since the stationary spectrum is bounded, 
the density of states scales linearly with the system size. 
It follows that in the basis of the stationary Hamiltonian the distance between the resonantly hybridizing states will also scale as $\sim N$, and the normalized effective width of the resulting Floquet states remains finite as $N\rightarrow\infty$, in contrast to the direct basis. In result, the numerically 
observed de-localization in the stationary basis, $L\sim 1$, 
persists with the up-scaling of $N$, while the normalized participation numbers and compactness decay, $P\sim 1/N$, $\zeta\sim1/N^2$; see  Fig.\ref{fig:2}(e).


Now we evaluate the structure of Floquet states in the eigenbasis of the stationary system. We calculate the period-averaged 
norm of each element of the Floquet state, $\|F_k^{(l)}\|=\frac{1}{T}\int_0^T |F_k^{(l)}(t)|^2 dt$, and its average energy, $E_l=\sum_k\lambda_k\|F_k^{(l)}\|$. We define the probability density function 
\begin{equation}
\label{eq:5}
\rho(\lambda,E)=\lim\limits_{\Delta\lambda,\Delta E\longrightarrow0}\frac{\langle\|F_k^{(l)}\|\rangle_{\lambda_k\in[\lambda,\lambda+\Delta\lambda], E_l\in[E,E+\Delta E]}}{\Delta \lambda\Delta E}, 
\end{equation}
which, additionally averaged over disorder realizations, 
essentially gives the contribution (weight) of stationary eigenstates with the energy $\lambda$ to (in) the Floquet states with the average energy $E$.

Figures \ref{fig:1}(a)-(c) illustrate the modification in the Floquet states by slow modulations  of the frequency $\omega=0.5$ (which is smaller than the characteristic spectrum width). 
For weak driving, $A=0.01$, the Floquet states are only slightly perturbed stationary eigen-states, 
with tangible distortions appearing, however, when the near-resonance condition holds, see Fig.~\ref{fig:1}(a). 
The main effect is observed for the quasi-extended states. Stronger modualtions, $A=0.1$, induce a progressive resonant hybridization of all states, 
including the strongly localized ones, see Fig.~\ref{fig:1}(b). As the driving amplitude grows  to $A=1.0$, the Floquet states become smeared over the stationary basis 
and completely loose their initial forms, see Fig.~\ref{fig:1}(c).

As de-localization of Floquet states in the basis of the stationary Hamiltonian 
is associated with thermalization of the system \cite{Lazarides2014,Rigol2014,Lazarides2015},
it is tempting to search for  it in our model. 
Usually, one checks if expectation values of certain observables are almost independent of
a particular type of the Floquet state, 
so we choose to follow the period-averaged energy, $E_l$. 
The results presented with  Fig.\ref{fig:1}(d) show the  development of a pronounced peak in distribution about $E=1.25$. Therefore, a progressively greater part of Floquet states assumes the average particle energy about the band center, which is a trait of the thermalization.

In conclusion, we studied the fate of the single-particle localization in a periodically-driven 1D lattice with a correlated disorder potential (which is the most conventional 
model of speckled optical lattices \cite{Modugno2006}). 
We demonstrate that the existence of the quasi-extended states,  appearing due to the presence of spatial correlations in the potential,   drastically changes
localization properties of the Floquet states. 
In the direct space we observe a trend to the homogeneity; namely, the 
decrease of localization for the strongly localized states and the enhancement of localization for the quasi-extended states, 
which relies on the resonant hybridization between the states from different bands. 
Ultimately, in the limit  of strong driving,  
most of the Floquet states completely loose their original properties 
and become  weakly localized  in the direct space. 
Moreover, they also de-localize in the basis of the stationary Hamiltonian and exhibit signatures of the thermalization. 
Our results can be straightforwardly generalized to the case of non-harmonic periodic driving \cite{Flach2016,Flach2014}. 
Finally, although the mobility edge in the used model is only an effective one (so that the quasi-extended states are not completely de-localized), 
our finding constitutes a step towards a much debated problem of the phenomenon of driven many-body localization in the presence of mobility edge \cite{Molina2009,Abanin2015,Lazarides2015}. 

The authors acknowledge support of Russian Science Foundation grant No.\ 15-12-20029 (analytical part) 
and Ministry of Education and Science of the Russian Federation Research Assignment No. 1.115.2014/K (computational studies). 
Numerical simulations were performed on the Lobachevsky University and Moscow State University Supercomputers and in the Joint Supercomputing Center of RAS.

\end{document}